\documentclass[12pt]{spieman}  
\usepackage{amsmath,amsfonts,amssymb}
\usepackage{graphicx}
\usepackage{setspace}
\usepackage{tocloft}

\newcommand{\lya}{Ly$\alpha$}
\newcommand{\RE}{$R_{E}$}

\title{The EarthASAP mission concept for a Lunar orbiting cubesat}

\author[a,*]{Ana I. G\'omez de Castro}
\author[a]{Leire Beitia-Antero}
\author[b]{Carlos  E. Miravet-Fuster}
\author[b]{L. Tarabini}
\author[b]{Albert Tom\'as}
\author[a]{Juan Carlos Vallejo}
\author[a]{Ada Canet}
\author[c]{Mikhail Sachkov}
\author[d]{Shingo Kameda}

\affil[a]{Universidad Complutense de Madrid, AEGORA Research Group, Fac. of Mathematics, Plaza de Ciencias 3, Madrid, Spain, 28040}
\affil[b]{SENER, Ingeniería y Sistemas, Carrer Creu Casas i Sicart, Parc de l'Alba, 86-88, Cerdanyola del Valles, Spain, 08290 }
\affil[c]{ Institute of Astronomy of the Russian Academy of Sciences, Pyatnitskaya st., 48, Moscow, Russia, 119017 }
\affil[d]{Department of Physics, Rikkyo University, Japan}

\cftpagenumbersoff{figure}
\cftpagenumbersoff{table} 
\begin{document} 
\maketitle

\begin{abstract}
There is  a growing interest in Lunar exploration fed by the perception that the Moon can be made accessible
to low-cost missions in the next decade.  The on-going projects to set a communications relay 
in Lunar orbit and a deep space Gateway, as well as the spreading of commercial-of-the shelf (COTS) 
technology for small space platforms such as the cubesats contribute to this perception. 
Small, cubesat size satellites orbiting the Moon offer ample opportunities to study the Moon and enjoy an advantage 
point to monitor the Solar System and the large scale interaction between the Earth and the solar wind. 
In this article, we describe the technical characteristics of a 12U cubesat to be set in polar Lunar orbit for this purpose
and the science behind it. The mission is named EarthASAP (Earth AS An exoPlanet) and was submitted to the Lunar Cubesats for 
Exploration (LUCE) call in 2016. EarthASAP was designed to monitor  hydrated rock reservoirs in the Lunar poles 
and to study the interaction between the large Earth's exosphere and the solar wind in preparation for future 
exoplanetary missions. 
\end{abstract}
\keywords{space vehicles; space vehicles: instruments; Moon; Earth; meteorites, meteors, meteoroids; comets: general}

{\noindent \footnotesize\textbf{*}Ana I. G\'omez de Castro,  \linkable{aig@ucm.es} }

\begin{spacing}{2}   

\section{Introduction}
\label{sect:intro}  
Cubesat technology started back in 1999. At that time, the California Polytechnic State University  and 
the Stanford University produced the basic CubeSat specifications as an attempt to facilitate the design, manufacture,
and testing of small satellites\footnote{www.cubesat.org}. Cubesats are modular structures with unit (1U) of dimensions
$10\times 10\times 10$ cm$^3$  and weight smaller than 1.5 kg that use commercial-of-the-shelf (COTS)
products. Usually, they are stacked in 2U, 3U up to 20U platforms originally intended  for low Earth orbit
(LEO) where the electronic components are not submitted to the hard environment of interplanetary space.

The increasing interest on Lunar exploration and the on-going initiative to build the Deep Space 
Gateway\footnote{exploration.esa.int/moon/59374-overview/} has open the path to the development of 
lunar cubesats. This new generation will need to confront important challenges such as the adaptation to 
interplanetary space or the development of communication strategies based on shared communications relays.
Moon's orbit offers ample opportunities to study the Moon and enjoys an advantage 
point to monitor the the Solar System and the large scale interaction between the Earth and the solar wind. 

In recognition of this interest, the European Space Agency (ESA) focused the 2016 SysNova call on lunar cubesats (LUCE);
ESA's SysNova calls open regularly\footnote{gsp.esa.int/sysnova} aiming at  getting innovative proposals in space technology.
LUCE Sysnova technical challenge was intended to address  four different topics: lunar resource prospecting,
environment and effects, science from or in the Moon and lunar exploration technology and operations demonstration.
The first version of the project we present in this article was submitted to this call, though it was not selected it evolved and matured 
to address some critical challenges such as the required instrument miniaturization, the  propulsion for orbit maintenance
or the on-board autonomy assuming sparse communication windows to an international communication relay. The project was named
Earth AS An exo-Planet (EarthASAP) as one of its key scientific objectives is using the lunar orbit vantage point to monitor the interaction between the 
Earth's exosphere and magnetosphere with the solar wind; also in preparation for the coming exoplanetary research missions.

In the following sections the scientific objectives (section 2), the mission concept (section 3), the mission and data handling 
(section 4) are described.  The article concludes with a brief summary together with the risk evaluation and the palliative solutions proposed by the team  (section 5).

\section{Scientific objectives}\label{sec:scientific_goals}
EarthASAP is intended to be in lunar polar orbit and run three research programs: the observation of the Earth's
magnetosphere and exosphere, the detection of diffuse matter within the
heliosphere (comets, asteroids, and dusty clouds) and the monitoring of the water content
and space weather in the lunar poles.

\subsection{Earth interaction with the solar wind}\label{sec:exoearth}

There is increasing evidence of the coupling between the upper atmospheric
layers (upper thermosphere) to the plasmasphere, the magnetosphere and to space weather conditions\cite{Qin2016}. At the base of the exosphere,  a population of energetic hydrogen atoms  
is generated as a result of this coupling. This layer is critical to understand the Earth's thermal flow and
particle escape, {\it i.e.} the planetary wind, as well as its relation  with solar activity. \par

\begin{figure}
\begin{center}
\begin{tabular}{cc}
\begin{tabular}{c}
\includegraphics[width=0.5\linewidth]{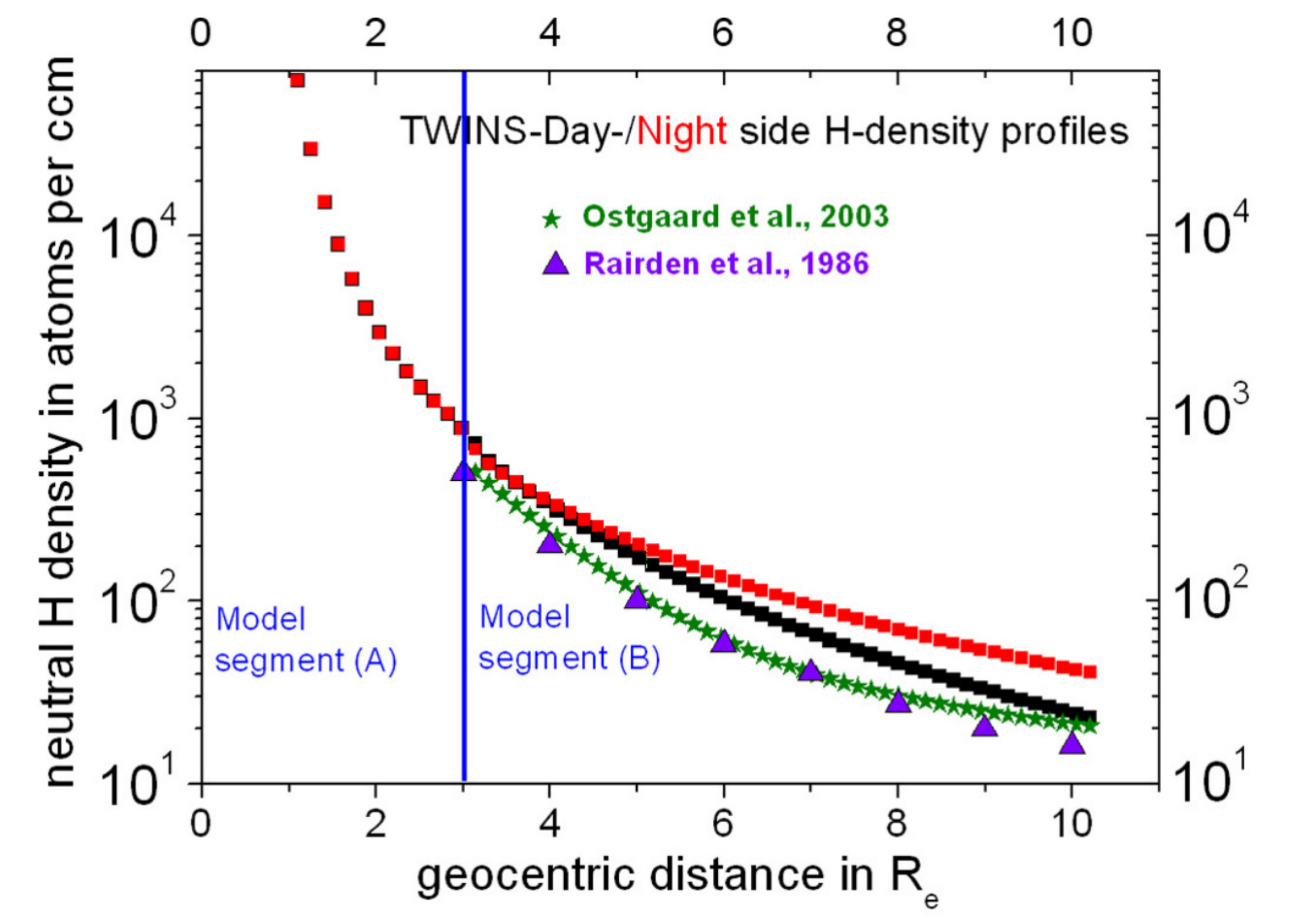} 
\end{tabular} &
\begin{tabular}{c}
\includegraphics[width=0.55\linewidth]{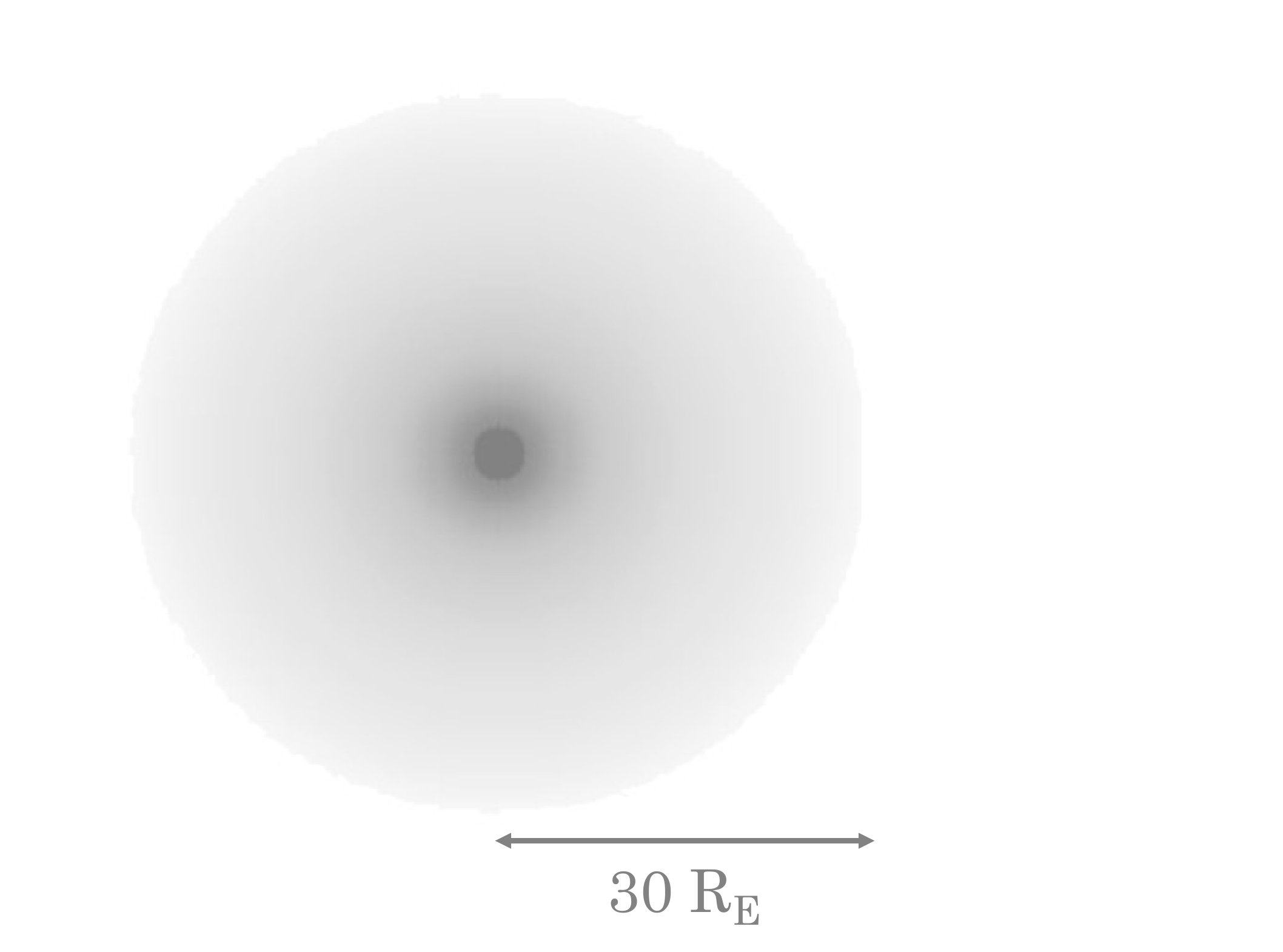} 
\end{tabular} \\
\end{tabular}
\caption{  {\it Left:} Radial distribution of hydrogen in the Earth's exosphere as derived from 
the  \lya\ measurements obtained by the TWINS mission \cite{Zoennchen2011}.
{\it Right:} Transmissivity of the Earth (including exosphere) to background \lya\ radiation; the radius is 30\RE \cite{GdC2018}
(black: zero transmissivity, white: full transmissivity, contrast in logarithmic scale).}
\label{fig:hydrogen_distrib}
\end{center}
\end{figure}

The most sensitive tracer to measure the distribution of exospheric hydrogen is the Lyman~$\alpha$ (Ly$\alpha$) 
transition.  Space probes such as the \textit{Imager for Magnetopause-to-Aurora Global Exploration} (IMAGE\cite{Fuselier2000}) or the \textit{Two Wide-angle Imaging Neutral-atom Spectrometers} (TWINS\cite{McComas2009})
mission have used \lya\ to derive the distribution of hydrogen  while navigating  within  the exosphere. 
The GEO instrument\cite{Mende2000} on board IMAGE  found that the particle
density does not decrease exponentially  with the distance to the Earth as predicted  
by the classical models\cite{Chamberlain1963}.  In fact, the density distribution is bimodal with a dominant
component peaking at 1\RE   ~ and an extended component reaching $n_{H} = 20$cm$^{-3}$ at
8\RE\cite{Ostgaard2003}. This result was further confirmed by the \textit{Ultraviolet Imaging
Spectrograph} (UVIS\cite{Esposito2004}) experiment on board the Cassini spacecraft.  
During Cassini's Earth swing by, the solar \lya~radiation scattered by the neutral hydrogen atoms of the
geocorona detected an extended exosphere around the Earth\cite{Werner2004}. Later
on, TWINS obtained the 3D distribution of \lya~emission within 8\RE~(and above 3\RE) around
the Earth and measured even higher density values \cite{Zoennchen2011}
(see Fig. \ref{fig:hydrogen_distrib}).
The recent images obtained by the Japanese Space Agency (JAXA) with the PROCYON probe
show that the Earth's exosphere extends further than 38\RE \cite{Kameda2017}.
Extended exospheres have also been detected  in Venus and Mars \cite{Shizgal1996} and this seems to be a common characteristic of terrestrial planets. 

Both IMAGE/GEO and TWINS reported variations in  the hydrogen distribution  by about
$\pm 20\%$ with time that apparently are not controlled by any solar quantity or geomagnetic 
parameter alone \cite{Bailey2013}. {\it A systematic monitoring of the Earth's hydrogen envelope and escape
is necessary to determine the processes controlling this evolution}.   The exosphere is 
mainly composed by neutral hydrogen that interacts very weakly with the diffuse solar wind
and the Earth's magnetic field. The exospheric ionization fraction is mainly controlled by the  
charge-exchange reactions between the neutral exospheric atoms and the fast charged 
particles accelerated by the geomagnetic field. The higher the ionization fraction, the more efficient 
is the solar wind in the removal of exospheric gas.  A systematic long-term monitoring of the Earth's 
hydrogen envelope and escape from outside of geocorona is necessary to determine the processes 
acting on the response of the exosphere to geostorms.

Missions like IMAGE or TWINS have been orbiting the Earth within the magnetospheric cavity 
(the cavity size is about 70,000~km towards the Sun and extends as far as 1000\RE\ towards the anti Sun, where the magnetotail
is located). As a result, the interpretation of the data has been hampered
by the difficulties to determine the \lya\ emission within a complex geometry
and to measure at the same time the solar flux and the heliospheric background. This
issue is critical since the exospheric emission at large \RE\ drops below the background
\lya\ radiation. As an example, at 10\RE\ the ratio of exospheric to
background \lya\ radiation ranges between 0.21 and 1.17 depending on the
strength of the variable background (estimate based on data from
IMAGE \cite{Ostgaard2003}). The images obtained by the Lyman Alpha Imaging Camera (LAICA) on-board
the PROCYON satellite show the potentials of wide field imaging from large distances
to capture the faint exospheric emission above the heliospheric background. \par

EarthASAP is designed to map the exosphere from outside, measuring simultaneously the exospheric
emission and the variable background obtaining very accurate measurements
of the hydrogen distribution at large Earth radii. This is feasible at a low
cost by using the Moon as a stable gravitational platform to orbit the Earth. A 3D image
of the Earth's exosphere and magnetosphere can be constructed  in one synodic month.   

EarthASAP's key objective  is  measuring the \lya~ source function
at scales as small as 0.1\RE\ (3~arcmin) 
and to resolve the contributions to the line excitation from solar radiation and other processes, such as collisions with 
high energy particles from the Sun or from internal magnetospheric sources, including the radiation belts.
Moreover, EarthASAP is designed to map the exosphere/magnetosphere radiation produced by neutral oxygen (OI)
and  ionized helium (HeII, n=2-3 resonance transition at 164 nm). Oxygen is expected to be abundant in 
Earth-like planets but even for the Earth, its exospheric distribution is poorly known at large Earth radii;
Oxygen is also the most highly referenced astronomical bio-signature gas.  All
the target tracers for the EarthASAP mission, HI, OI and HeII, are also important
tracers of energetic neutral atoms (ENAs). ENAs are created in charge-exchange
collisions between hot plasma ions (protons, alpha particles) and the
cold neutral gas in the  magnetospheric/exospheric environment. Therefore, the
spatial distribution of ENAs emission indicates the location of frequent
collisions  with fast particles (solar wind, magnetic reconnection regions,
radiation belts) and can be used as a tracer of these phenomena. The wide field
imaging and the rapid read-out of EarthASAP is designed to enable monitoring 
variable phenomena and isolate their contribution to the total radiative budget.

\subsubsection{Connection with exoplanetary research}

The detection of exoplanets has opened the possibility of studying planetary exospheres and 
atmospheric escape in planetary systems  submitted to very different stellar radiation fields and space weather conditions.
The transit of planetary exospheres in front of the stellar disk produces a net absorption that has been detected 
in the  stellar Ly$\alpha$  profile of large, massive planets in close orbit around their parent stars:
HD 209458b \cite{VidalMadjar2003}, HD 189733b \cite{Lecavelier2010} and
GJ 436b \cite{Kulow2014,Ehrenreich2006}. The next step is measuring atmospheric escape 
in Earth-like exoplanets. The first evaluations are promising, especially for planets orbiting in the habitable zone 
around M-type stars \cite{GdC2018}. The high optical depth
of the \lya\ line results in very low gas columns ($N_{H} > 5\times 10^{17}$ cm$^{-2}$) being sufficient to block
the \lya\ radiation from the star without producing noticeable effects in the rest of the stellar
spectral tracers \cite{GdC2016}. The radius at which \lya\ becomes
optically thick, $\tau_\text{\lya} > 1$, sets the geometric cross section of the planet to
the \lya\ radiation which is significantly larger than the planet itself. As a result, the transit duration 
and the shape of the \lya\ light curve depend strongly on exospheric properties. 
Observations of exoplanetary \lya ~ transits will enable understanding the 
interaction of the Earth's atmosphere with the Sun at the time when life emerged on Earth, 
an uncertain date between 2.7 and 4.3 Gyr ago \cite{Buick2007}.  In preparation for these 
activities, a better understanding of the Earth's environment is required.

\begin{figure}
  \centering
  \includegraphics[ width = 0.4\textwidth]{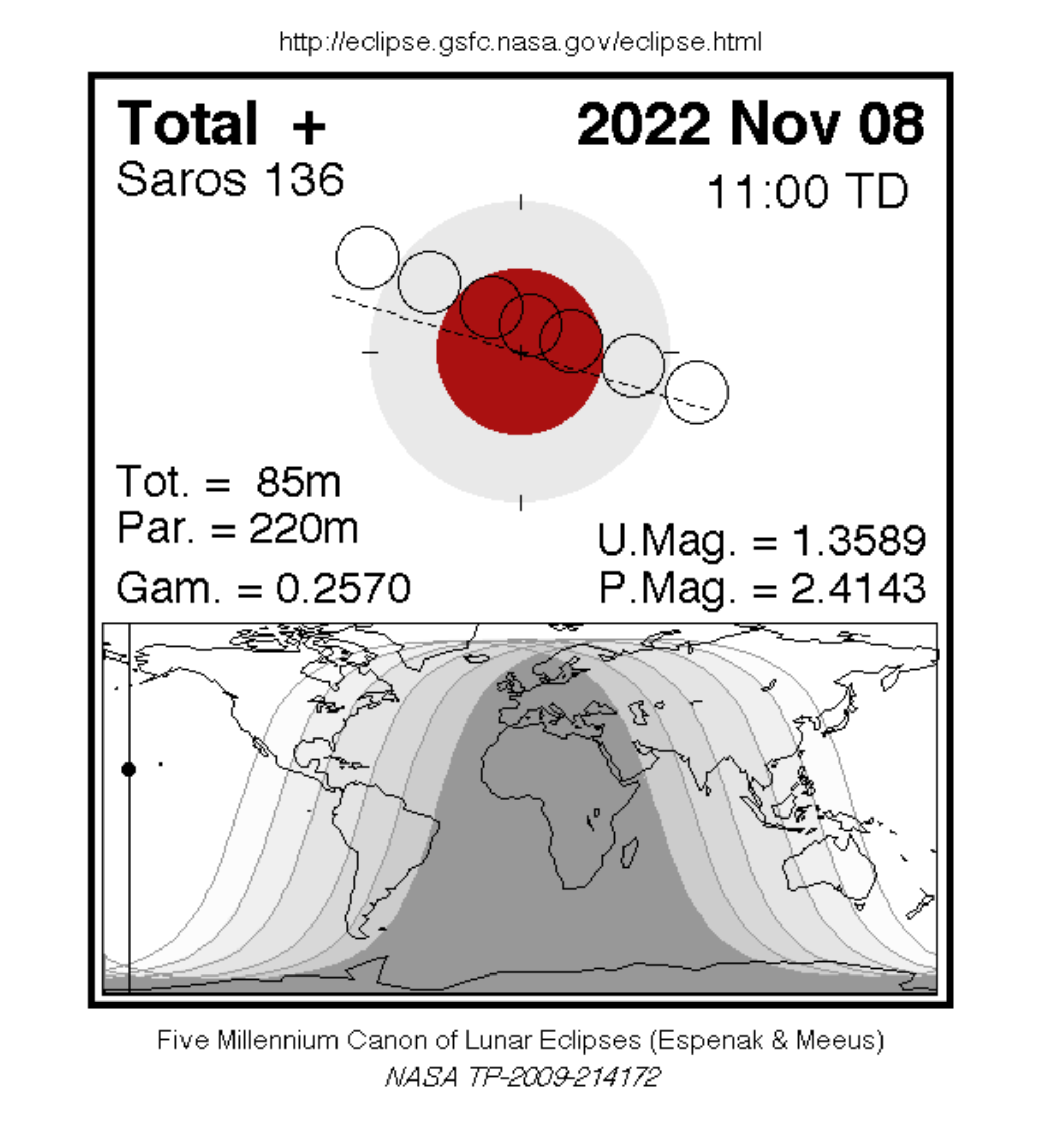}  
  \includegraphics[ width = 0.4\textwidth] {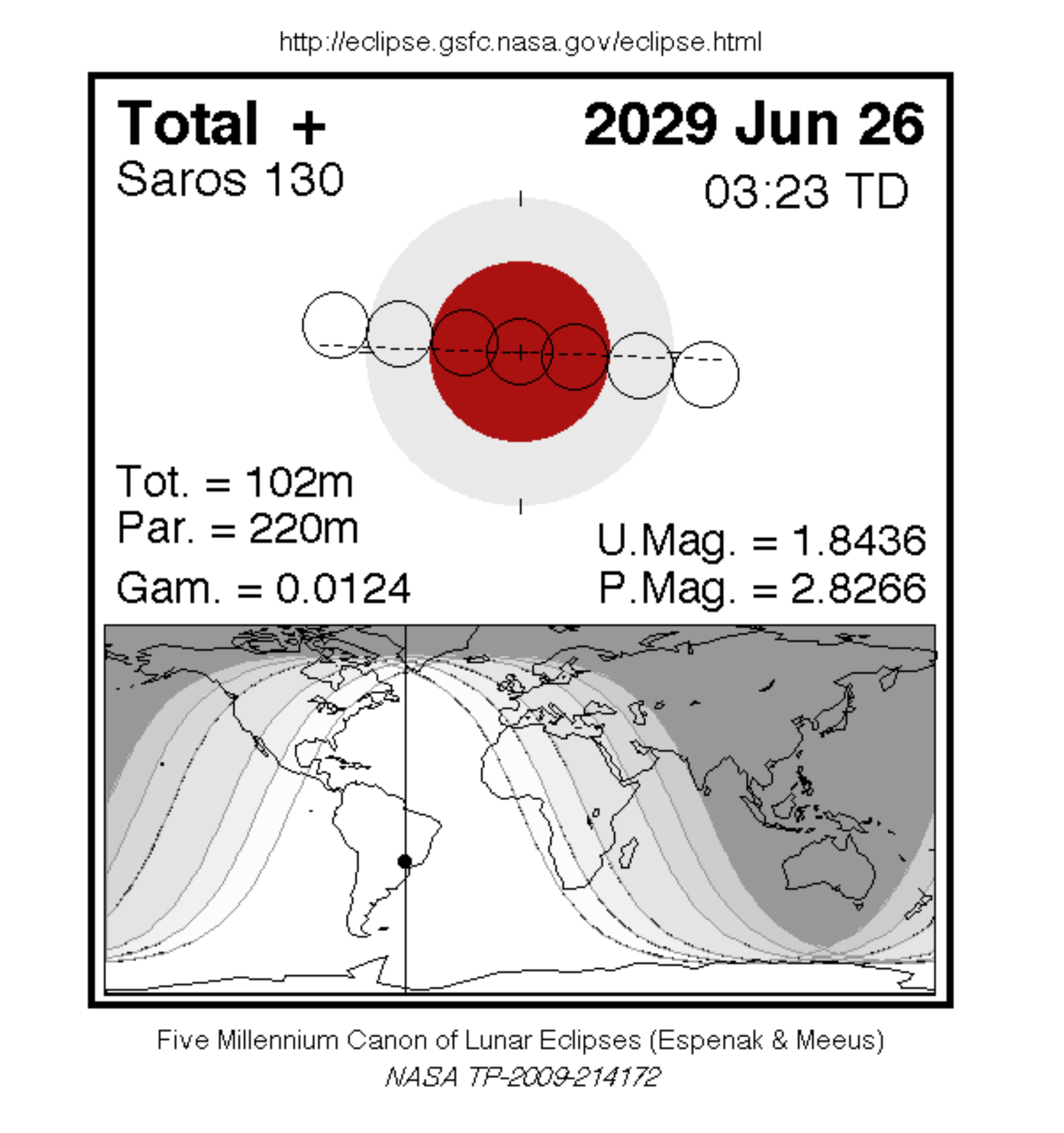} \\
  \caption{ Earth-Moon configuration during the next total lunar eclipses: November $8^{th}$, 2022 and June 26th, 2029. 
  At central eclipse, the Earth, the Moon and the Sun are nearly aligned enabling  a unique 
  observation of the Earth's geocorona from Lunar orbit.}
  \label{lunar_eclipse}
\end{figure}

\subsubsection{Earth observations during lunar total eclipses}\label{sec:lunar_eclipse}
During the next lunar eclipses of  November $8^{th}$, 2022 and June 26th, 2029, there will be a unique chance 
to observe the Earth's exosphere while transiting the solar disk (see Fig. \ref{lunar_eclipse}). 
This viewing angle will provide exclusive information on the scattering and absorption of solar
photons by the uppermost layers of the Earth's atmosphere to compare with the
theoretical predictions of the \lya\ light curves.

\subsection{Diffuse matter within the heliosphere: comets, asteroids, and dusty clouds}\label{sec:diffuse_matter}

The resonant scattering of solar \lya~ photons by neutral hydrogen within the
heliosphere produces a bright far ultraviolet background first reported
from OGO-5 data \cite{Bertaux1971,Thomas1971}. The slow flow of hydrogen atoms comes from
the Local Interstellar Cloud (LIC); as the Solar System crosses the LIC, the heliosphere
creates a shock front that only neutral particles can penetrate. The shock front
is at 94 AU from the Sun \cite{Stone2005} in the upwind direction, ecliptic
longitude (J2000) $225^o.4 \pm 0.^o5$ and latitude $5^o.1\pm0.^o2$ \cite{Witte1996}.
The heliospheric \lya\ intensity varies as the Sun rotates,
especially at low latitudes since the Sun's ultraviolet emission is enhanced in
localized active regions; however variations occur at a slow pace \cite{Rairden1986}. \par

Surveying systematically the heliosphere in \lya~ enables the detection and
study of  comets photoevaporation. \lya~ observations
are an ideal tool for searching for comets since neutral hydrogen is the main component of
the coma. These observations can be used to derive the water production rate of
the comet as a function of time \cite{Bertaux1998,Makinen2000}. EarthASAP is designed to grow on the
experience of SOHO/SWAN providing higher angular resolution ($0.05^o$ instead
of $1^o$) and a wide field of view to study the detailed structure of the
evaporation process and the interaction of the coma with the heliospheric
magnetic field. This extra angular resolution will result in an
  enhanced detectability
since cometary tails are clumpy structures. High resolution imaging
of hydrogen distribution in the comet envelop and tail is a very sensitive probe to study comets'
photoevaporation and ionization by solar photons. Moreover, the tail  interacts with the solar wind
through magnetic processes and charge-exchange reactions; hence, UV observations are very useful
to study the solar wind properties along the path of the comet \cite{Shustov2018}. EarthASAP angular resolution is designed to be similar to that of ALICE\cite{Stern2007}. \par

The far ultraviolet range is very sensitive to small dust columns. As a result, dust clouds in the Earth-Moon environment 
can be monitored such as those located in the libration points or those associated with small bodies streams
such as the Orionids, Taurids or Leonids orbits.

\begin{figure}
  \centering
  \includegraphics[ width = 0.5\textwidth]{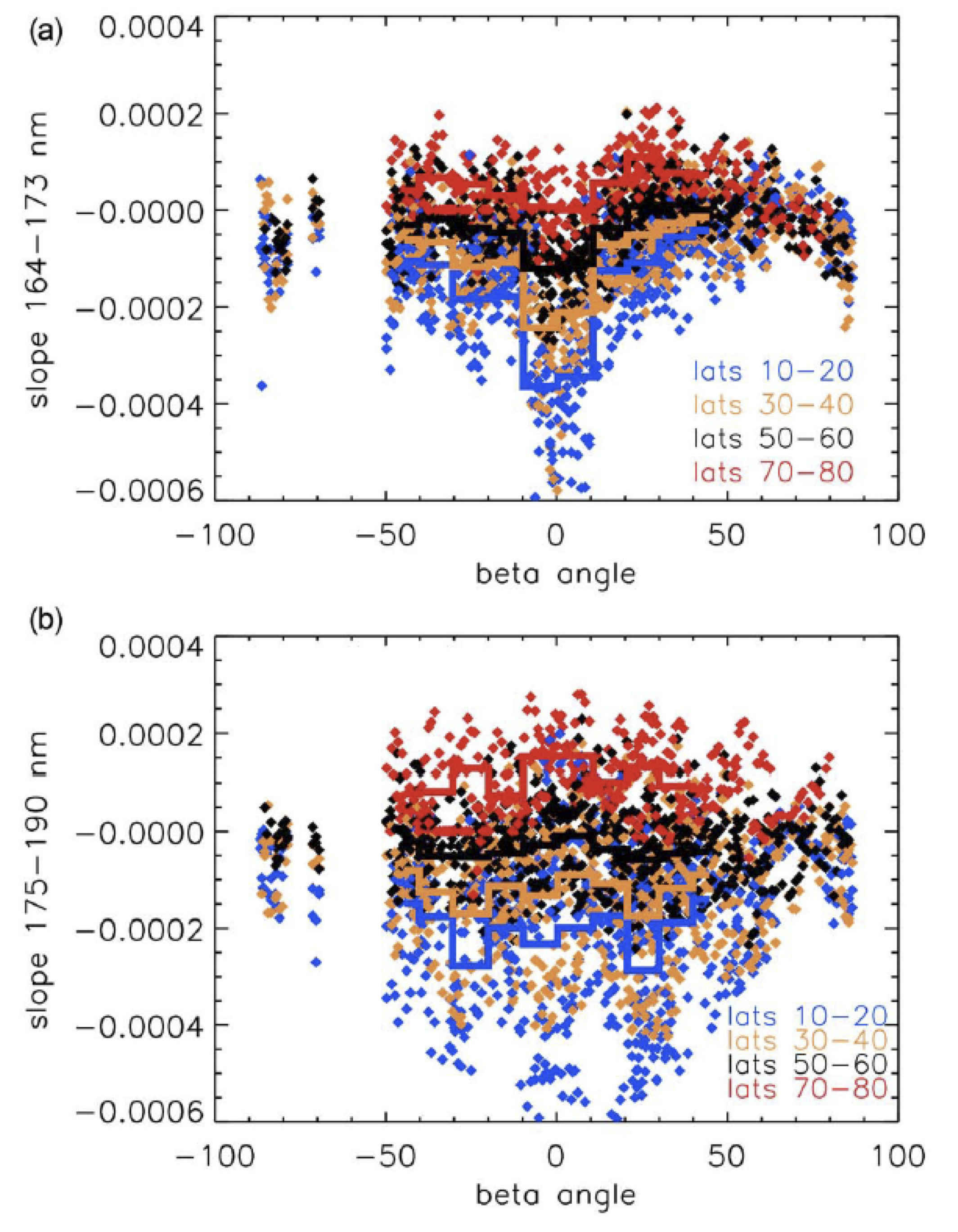}    
   \caption{Measurements of lunar hydration using LAMP 
\cite{Hendrix2012}. The  slope is plotted versus beta angle:
(a) 164-173 nm and (b) 175-190 nm. Negative beta angles
correspond to morning, positive beta angles correspond to
afternoon.  Average slopes in each 10$^o$ beta angle range 
are overploted as thick lines.}
  \label{fig:lunar_hydration}
\end{figure}

\subsection{Water content and space weather in the Moon poles}\label{sec:lunar_poles}

Future lunar settlements will be most likely be sited on the Moon poles. Thus, it is
crucial to determine the abundance of the water content  in hydrated
rocks. The instrument Lyman Alpha Mapping Project (LAMP) on board the Lunar Reconnaissance Orbiter \cite{Gladstone2012} 
mapped the hydration and weathering of the lunar terrain by measuring the slope
of the UV spectrum in 164-190 nm range \cite{Hendrix2012}. 
The presence of a strong water absorption edge in the far-UV (near 165
nm) enables the study of lunar hydration by a simple linear fitting  
of the  reflectance spectrum in the 164-173 nm range, where even the 
small water abundance affects significantly the slope. As shown in  Figure \ref{fig:lunar_hydration},
significant variations are detected with latitude; these variations are compared
with the slope in the  175-190 nm range that it is not affected by  hydration effects. 
In fact, the analysis of LAMP dayside data shows that the 175-190 nm region 
is a good indicator of weathering and composition  \cite{Hendrix2016}. The steeper
(redder) slopes are consistent with increased hydration
earlier and later in the day, and at higher latitudes.

\begin{figure}
  \centering
  \includegraphics[ width = 0.9\textwidth]{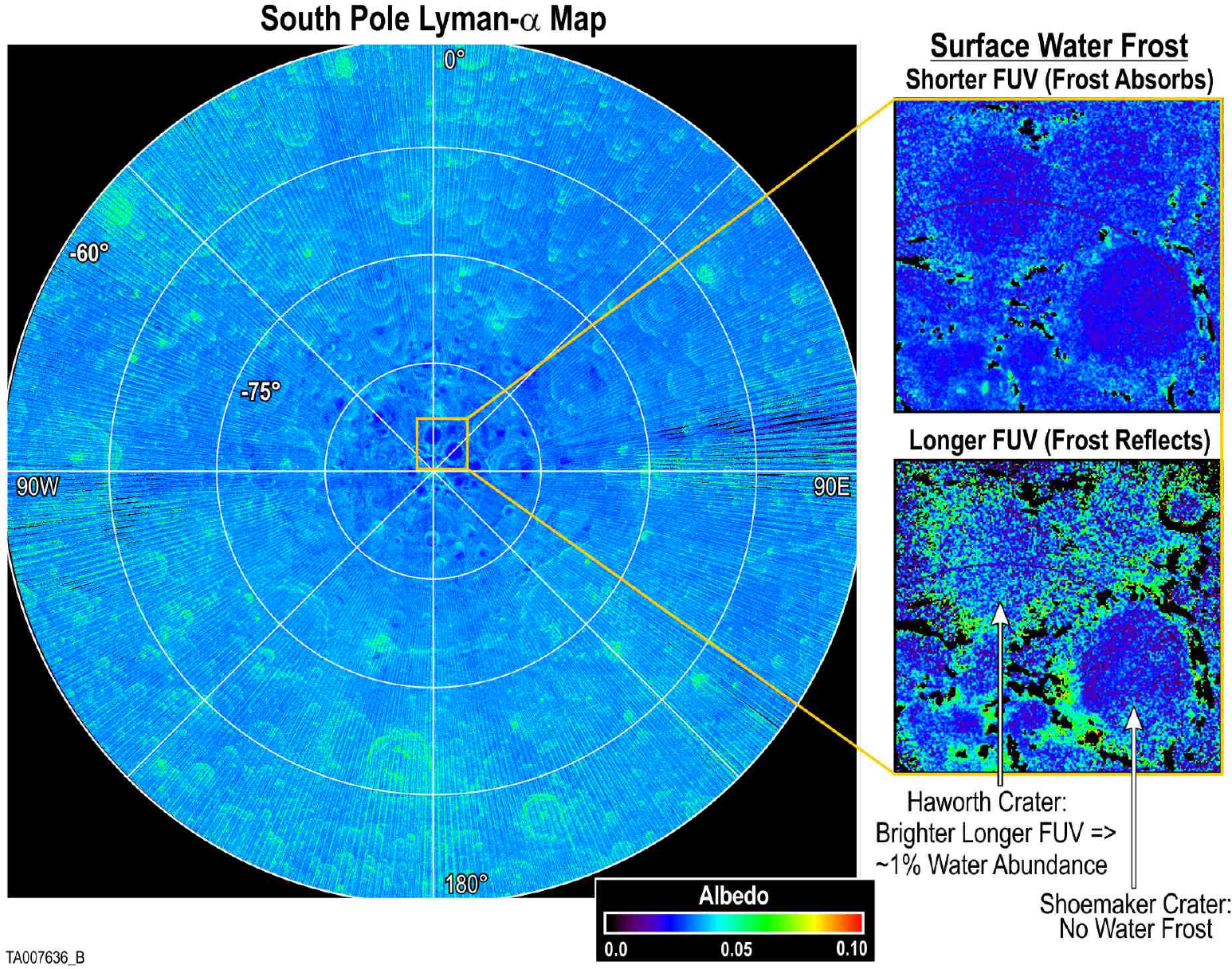}  
  \caption{Measuring the hydration of rocks on the lunar south pole from FUV imaging.
  [Courtesy from NASA, LRO/LAMP mission.]}
  \label{fig:frost}
\end{figure}

EarthASAP is designed to monitor the poles with a wide field of view 
(roughly 174km $\times$ 262km) per image
and a resolution of $\sim430$m. Though LRO/LAMP images have better resolution 
(down to 100m \cite{Chin2007}) EarthASAP's wide field of view makes feasible
differential measurements of the far UV emission variability and hence, of the
variations of surface ice and frost in the polar regions. In Figure \ref{fig:frost}, the procedure is illustrated from the
images obtained by LRO/LAMP of the south lunar pole.\par

\subsubsection{Clouds of dust in the Moon}

During the Apollo era, astronauts saw \textit{horizon glow} and \textit{streamers} in
the Moon's outmost atmosphere, or "exosphere". Since then, many scientists have suggested
that these phenomena were caused by sunlight scattered by dust grains in the
exosphere. Questions about how lunar dust and dusty plasmas are charged, mobilized
and transported remain at the centre of dusty plasma studies. EarthASAP is designed to
answer these questions by detecting the presence of dust clouds and plumes,
as well as their interaction with solar radiation and particles.

\section{Mission concept }\label{sec:mission_concept}

The EarthASAP mission covers several science themes devoted
  to the study of the Earth and the Moon, listed in Table \ref{tab:topics}. 
The previous section presented the desired science goals, that are achieved by 
different objectives, which in turn, impose the mission requirements.
In Table \ref{tab:topics} also the specific results to be obtained are
shown.

\begin{table}
\centering
\caption{\textbf{Scientific goals, requirements and results traceability matrix}}
\begin{tabular}{|l|l|l|}
\hline
Theme &
\multicolumn{2}{|l|}{Earth interaction with Solar Wind} \\ 
Goal & 
\multicolumn{2}{|l|}{To observe the Earth's magnetosphere and exosphere in UV} \\ 
\hline\hline
Specific Objectives & 
Requirements & 
Results \\
\hline
\begin{minipage}[t]{0.3\textwidth}
\begin{itemize}
\item Imaging capabilities in FUV range from $115$nm to $180$nm.
\item To fit the Earth's magnetosphere in one single image 
\item To resolve structures in the Earth's exosphere  $< 0.1$\RE
\item To observe the spatial distribution of ENAs 
\end{itemize}
\end{minipage}  &
\begin{minipage}[t]{0.3\textwidth}
\begin{itemize}
\item Solar blind detection system in the $115$nm to $180$nm range
\item Field of view  $22^o \times 20^o$ 
\item Angular resolution better than $3$ arcmin
\item FUV range, includes \lya (121.6nm), OI ($130.3$nm) and HeII ($164.0$nm). 
Narrow band filtering (FWHM=$5$nm) at these wavelengths is needed
\end{itemize}
\end{minipage} &
\begin{minipage}[t]{0.3\textwidth}
\begin{itemize}
\item Production of first $3$D map of Earth's exosphere 
\item Determination of the exospheric emission and transfer function of \lya\ photons
through the Earth's exosphere.
\end{itemize}
\end{minipage}  \\
\hline 
\hline

Theme &
\multicolumn{2}{|l|}{Diffuse matter within the heliosphere} \\ 
Goal & 
\multicolumn{2}{|l|}{To detect extended features against the heliospheric UV background} \\ 
\hline\hline
Specific Objectives & 
Requirements & 
Results \\
\hline
\begin{minipage}[t]{0.3\textwidth}
\begin{itemize}
\item Detect the $0.16$kR exospheric \lya\ emission at 15~\RE \;over the heliospheric background with SNR=$10$
\item To resolve structures in the Earth's exosphere  $< 0.1$\RE
\end{itemize}
\end{minipage} &
\begin{minipage}[t]{0.3\textwidth}
\begin{itemize}
\item (FoV and angular resolution as above)
\end{itemize}
\end{minipage}  &
\begin{minipage}[t]{0.3\textwidth}
\begin{itemize}
\item Derivation of the water production rate of a comet as function of time
\item Produce maps with new extended features such as 
comet tails, dust clouds and HI filaments
\item Planet $9$ might be detected if surrounded by a cloud of small bodies and dust from outer Solar System
\end{itemize}
\end{minipage}   \\
\hline 
\hline

\end{tabular}
\label{tab:topics}
\end{table}

\begin{table}
\centering
\caption{\textbf{(Cont.) Scientific goals, requirements and results traceability matrix}}
\begin{tabular}{|l|l|l|}
\hline
Theme &
\multicolumn{2}{|l|}{Water content on the Moon poles} \\ 
Goal & 
\multicolumn{2}{|l|}{To image lunar hydration, including FUV variability } \\ 
\hline\hline
Specific Objectives & 
Requirements & 
Results \\
\hline
\begin{minipage}[t]{0.3\textwidth}
\begin{itemize}
\item Image the lunar poles
\item Imaging spots of around $175$ km $\times$ $260$ km 
\item Resolve structures on the  lunar surface $< 0.5$ km size
\item Differential measurements, analysis of FUV slope
\end{itemize}
\end{minipage}  &
\begin{minipage}[t]{0.3\textwidth}
\begin{itemize}
\item Lunar polar orbit
\item (FoV and angular resolution as above)
\item Narrow band filtering at $170$nm, $175$nm, $187$nm and $190$nm
\end{itemize}
\end{minipage} &
\begin{minipage}[t]{0.3\textwidth}
\begin{itemize}
\item Time series of water contents on the Moon's poles
\end{itemize} 
\end{minipage}  \\
\hline 
\hline

Theme &
\multicolumn{2}{|l|}{Space weather on the Moon poles and lunar exosphere} \\ 
Goal & 
\multicolumn{2}{|l|}{To detect the presence of dust clouds and plumes in the lunar exosphere} \\ 
\hline\hline
Specific Objectives & 
Requirements & 
Results \\
\hline
\begin{minipage}[t]{0.3\textwidth}
\begin{itemize}
\item Detect and qualify dust clouds
\item Detect the 0.16kR exospheric \lya\ emission at 15~\RE~over the heliospheric background with SNR=$10$
\end{itemize}
\end{minipage}  &
\begin{minipage}[t]{0.3\textwidth}
\begin{itemize}
\item (FoV and angular resolution as above)
\item Step filter (f.i,, F$_2$Ba) to be used together with no-filter to measure extinction
\item Efficiency of the optical system and aperture
\end{itemize}
\end{minipage} &
\begin{minipage}[t]{0.3\textwidth}
\begin{itemize}
\item Confirm if sunlight scattered by dust grains produce the observed features such as horizon glows and streamers
\end{itemize} 
\end{minipage}  \\
\hline 
\hline

\end{tabular}
\end{table}

These scientific requirements can be fulfilled by the EarthASAP proposal, based on 12U cubesat, with 8U dedicated to the scientific
payload and 4U to the service module. 

\subsection{Scientific payload}

\begin{figure}
  \centering
  \includegraphics[width = \textwidth]{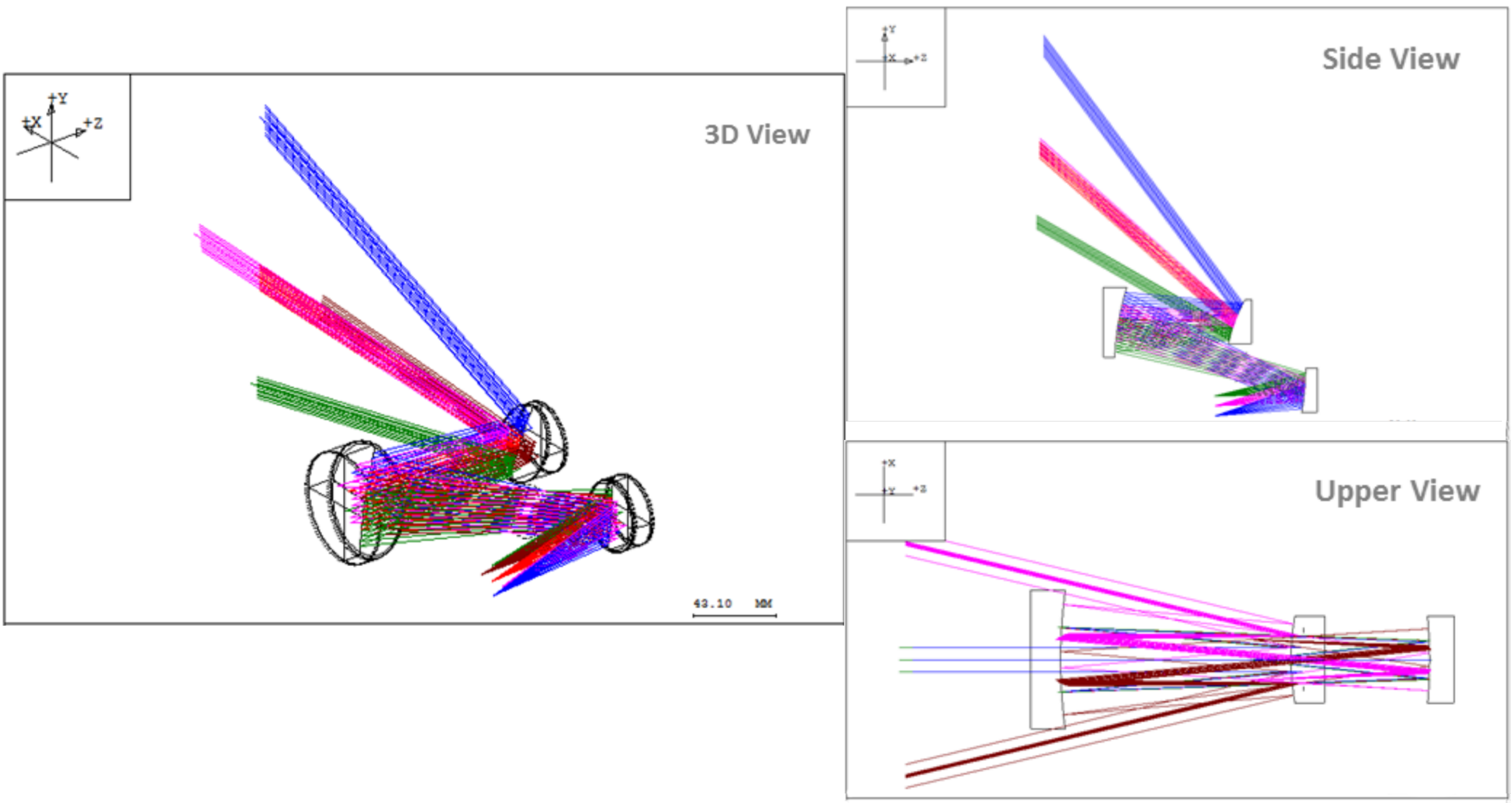}
  \caption{Optical layout for the WALRUS system implementation.}
  \label{fig:optical_layout}
\end{figure}

The optical design of the imager is an evolution of the Wide Angle Large Reflective Unobscured
System (WALRUS \cite{WALRUSpatent,WALRUSarticle}). It is a three-element
mirror system, with spherical primary and tertiary mirrors and a prolate secondary that
results in a system completely free of chromatic aberrations, with a
wide field of view ($22^o \times 20^o$), focal-ratio of $f/4.5$ and a
compact design ($65\times115\times190$mm$^{3}$) suitable for
a cubesat. It also provides an excellent image quality with flat field, covering
the spectral range 100-300 nm. The optical
layout is presented in Fig. \ref{fig:optical_layout} and the parameters of the proposed
designed are listed in Table \ref{tab:telescope_parameters}. Filters will be
  inserted in a filter wheel before the detector; based on previous developments
  \cite{Rodriguez2016,GdC2011,Fernandez2009} a transmittance of 7\% with 
high out-of-band rejection is expected in the narrow band line filters tuned to 
Ly$\alpha$, OI and HeII. 

The payload module is assumed to weight 8 kg, consume 4W and require a pointing
accuracy of $14''$ ($3\sigma$). Due to the dimensions of the telescope, the payload
module requires a 8U system including the detector, 
an MCP with CMOS read-out. The focal length and image size of the telescope are compatible
with the use of the multichannel plate PHOTEK image intensifier MCP118\footnote{\url{http://www.photek.com/pdf/datasheets/detectors/DS001_Image_Intensifiers.pdf}}, with an input window of MgF$_{2}$, a photo-cathode of CsI and a phosphor anode of P46/P47. The MCP118 image intensifier has a typical effective resolution of 40-50 lp/mm, leading
to an effective sampling size of around 15$\mu$m and resulting in 2.5 samples inside
the optics blur diameter of 0.04mm. We have checked that it is actually possible to
reduce that blur diameter for all points in the field of view to get 500 
effective samples across the detector size of 18mm. The  CMOS detector
is planned to be an AMS (CMOSIS) 5.5$\mu$m pixel size based on previous experience. \\

\begin{figure}
  \centering
  \includegraphics[width = \textwidth]{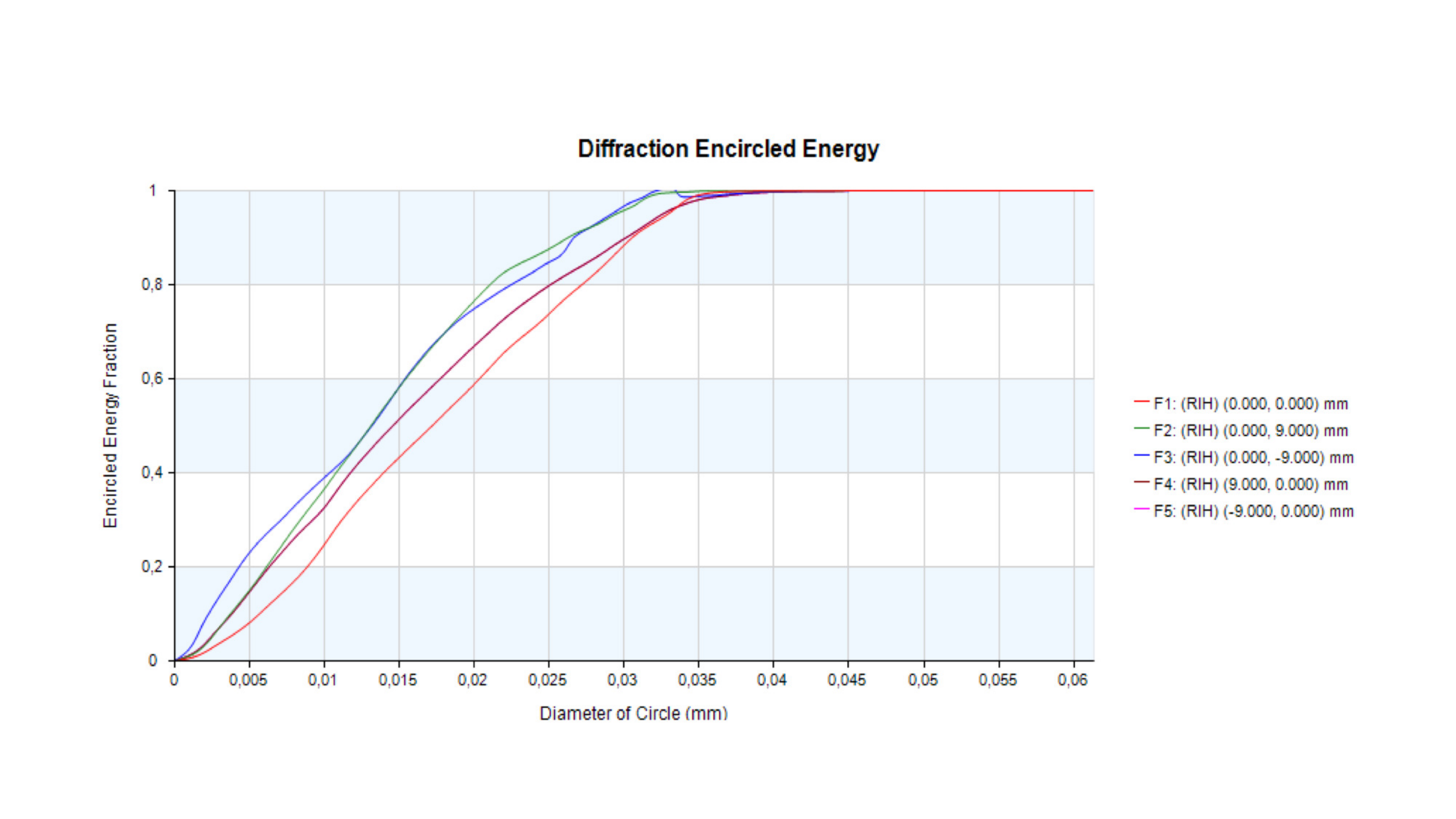}
  \caption{Encircled energy of the proposed design.}
  \label{fig:encircled_energy}
\end{figure}

\begin{figure}
  \centering
  \includegraphics[width = \textwidth]{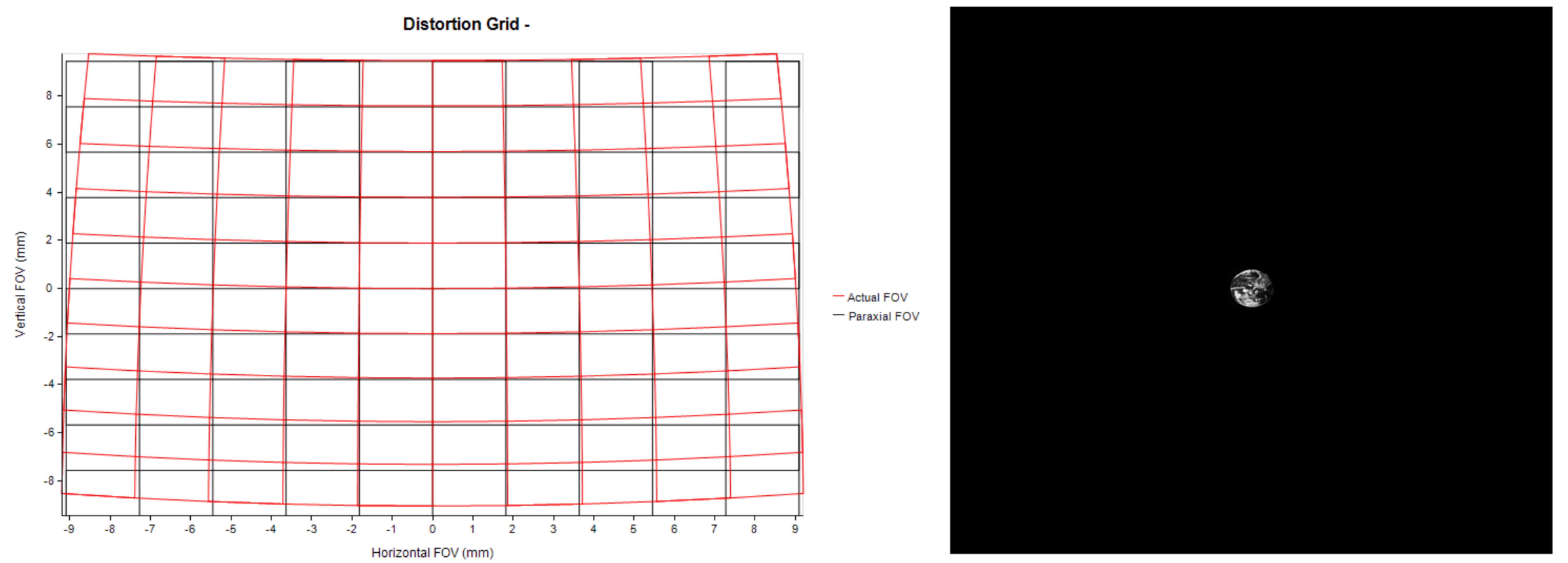}   
    \caption{{\it Left:} Optical distortion of the proposed design. {\it Right:} Simulation of the Earth image 
  as seen with the proposed design}
  \label{fig:distortion}
\end{figure}

The encircled energy of the proposed design is included in Figure \ref{fig:encircled_energy}. As shown, 
the preliminary optimization performed is enough to bring the optics blur diameter to a value lower 
than 0.04mm, for all points in the field of view. This will result in the optics providing around 500 effective samples across the detector size of 18mm.
The proposed optical design is subjected to a moderate amount of barrel distortion (peak value of -4.5\%), 
as can be observed in Figure \ref{fig:distortion}. This effect will be corrected for as part of the image post-processing and 
thus it is not considered relevant in terms of overall optical performance. The simulated effect of both the image 
quality and optical distortion of the proposed design on the Earth's image is also displayed in the figure.

Sensitivity requirements are met with this design and configuration. The \lya~ emission from the
Earth's exosphere can be mapped with  SNR$>$10 up to 15\RE~ from the Earth's centre. Under the
worst conditions (solar zenital angle $90^o$), the exosphere produces a \lya~ flux
of 0.16kR at 15\RE~ that is observed agains a 1.1kR heliospheric background
\cite{Pryor1998,Quemerais2014}. At the Moon's distance, the count rate from
15\RE~  is expected to be 115 counts s$^{-1}$ per angular resolution element ($0.05^o$ or 335km$^{2}$) 
at the telescope entry pupil. However, UV optics is not very efficient; the expected transmittance of the  optical 
system at \lya~ is about 2\% and the detector QE at this wavelength is ~30\%. As a result,  the count rate per angular resolution element on 
the detector is significantly lower (0.69 counts s$^{-1}$) and it will be spread in $\sim$6 pixels
to sample the PSF by $2.5 \times 2.5$~pix$^2$; SNR = 10 will be reached in less than
15 minutes with the optical elements and MCP detectors proposed for EarthASAP. 
The measurements will be made in TIME-TAG mode with the detector
being read every 40ms, therefore the total count rate will be 280,000, well within
the safety margins of conventional MCP detectors.  These calculations are made using state of the art 
technology; the baseline MCP detector is currently under development for the WSO-UV mission \cite{WSO2018}
and it is based on the heritage of the ISSIS instrument \cite{GdC2014} though its QE has been  enhanced
to reach the sensitivity of the  Hamamatsu detectors used in the  Lyman Alpha Imaging Camera (LAICA) 
on board the Japanese  micro-spacecraft PROCYON\footnote{Proximate Object Close Flyby with Optical Navigation} 
with a rather similar optical configuration \cite{Kameda2018}. Similar results have  also been obtained 
by SOHO/SWAN \cite{Baliukin2019}.

\subsection{Service module}

\begin{table}
\centering
\caption{Telescope parameters for the adapted WALRUS system. The stated dimensions correspond only to the presented optical layout, prior to baffling for stray-light control.}
\begin{tabular}{|c|c|}
\hline
\rule[-1ex]{0pt}{3.5ex}  Parameter & Value \\
\hline\hline
\rule[-1ex]{0pt}{3.5ex}    Type & All-reflective (WALRUS) \\    \hline
\rule[-1ex]{0pt}{3.5ex}    Focal Length & 48 mm \\    \hline
\rule[-1ex]{0pt}{3.5ex}    F-number & $f/4.5$ \\    \hline
\rule[-1ex]{0pt}{3.5ex}    Spectral range & 100-300nm \\    \hline
\rule[-1ex]{0pt}{3.5ex}    Image Size & 18mm \\    \hline
\rule[-1ex]{0pt}{3.5ex}    Field of View & ($22^o \times 20^o$) \\    \hline
\rule[-1ex]{0pt}{3.5ex}    Encircled energy diameter & $<0.05^o$ \\    \hline
\rule[-1ex]{0pt}{3.5ex}    Dimensions $(X\times Y\times Z)$ & $65\times115\times190$mm$^{3}$ \\    \hline
  \end{tabular}
  \label{tab:telescope_parameters}
\end{table}

\begin{figure}
\begin{center}
\begin{tabular}{cc}
\includegraphics[width=0.5\linewidth]{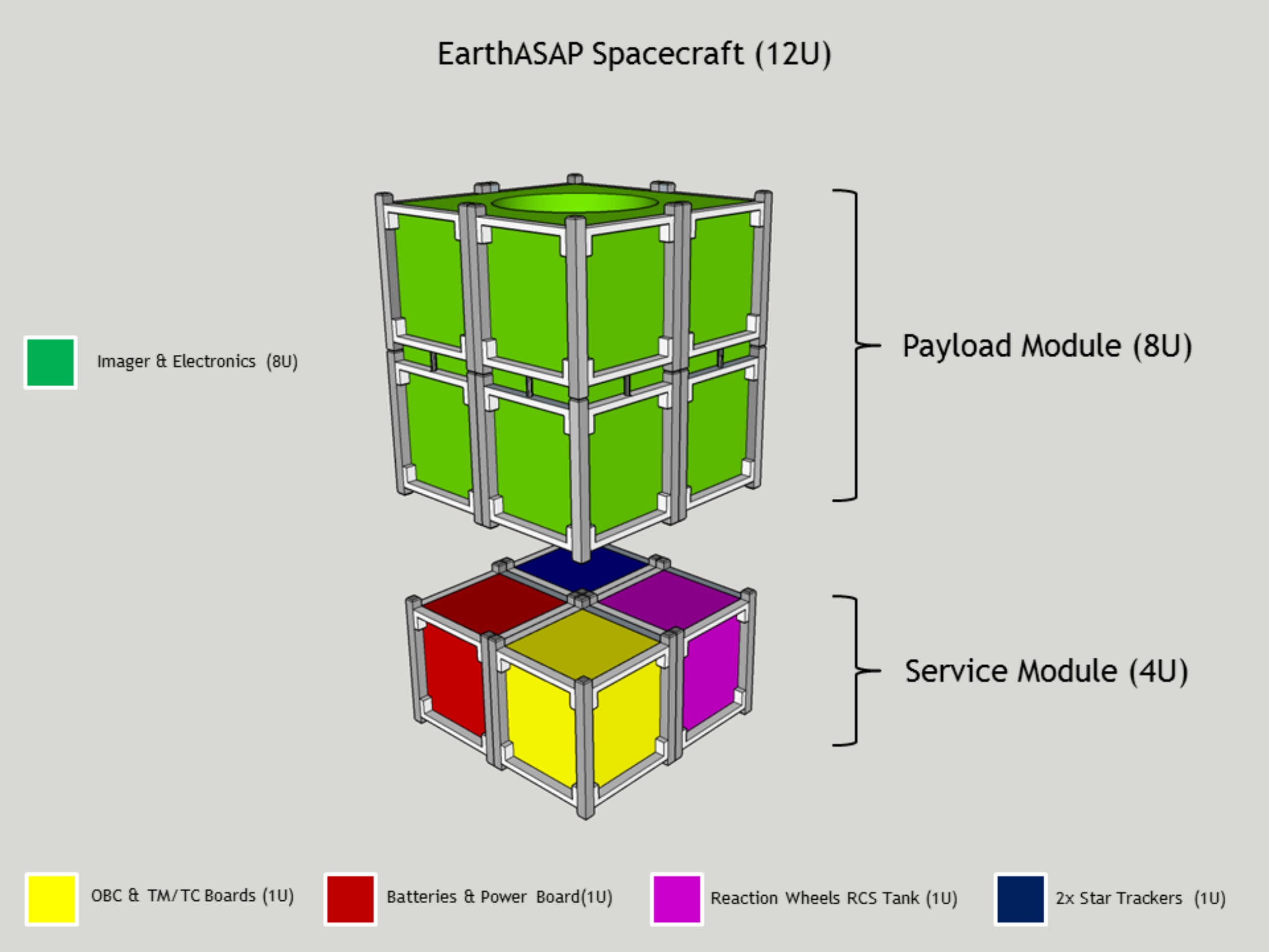} &
\includegraphics[width=0.5\linewidth]{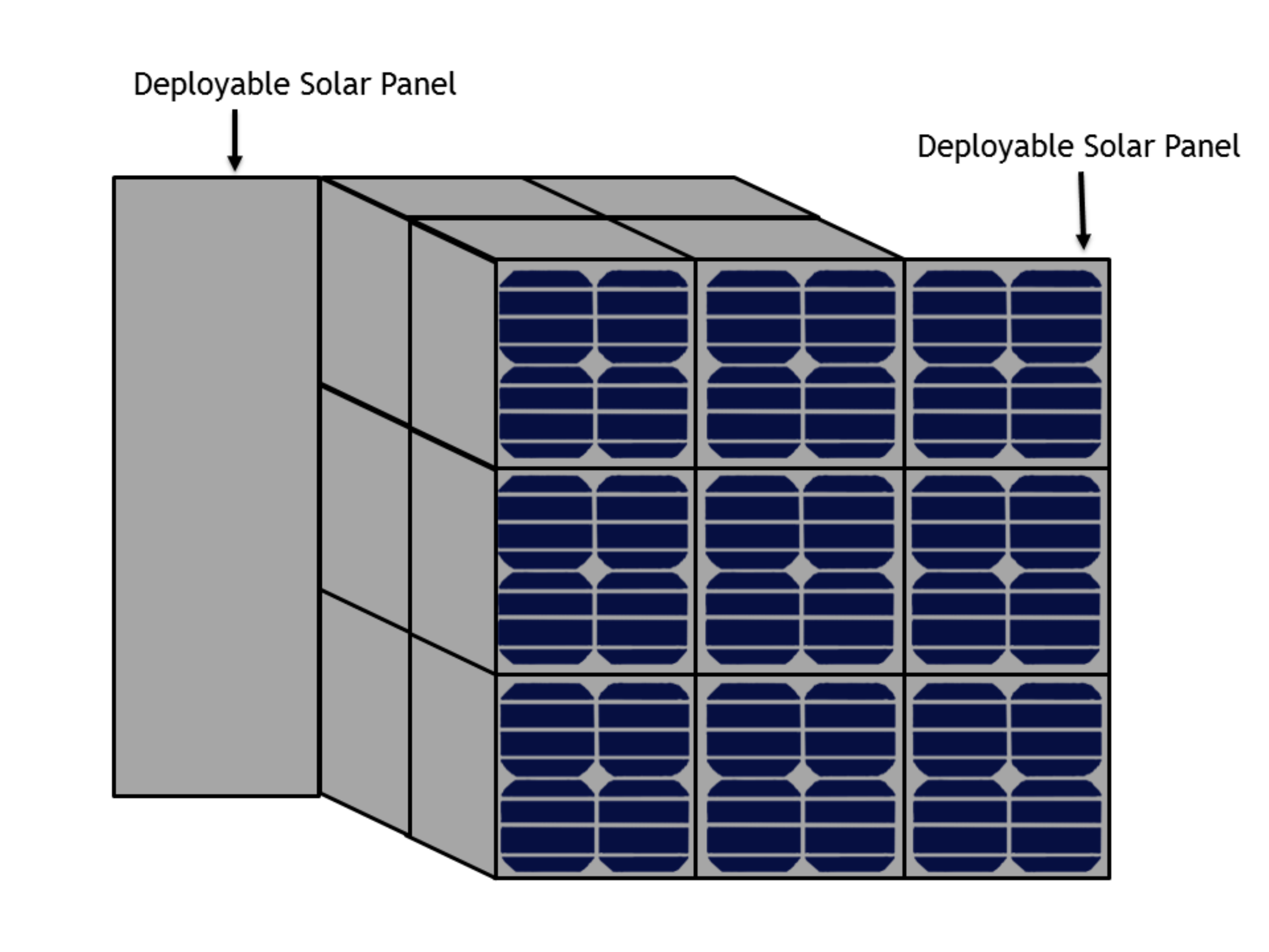} \\
\end{tabular}
\end{center}
\caption{
Integration of the proposed elements of the EarthASAP mission in a $12$U cubesat structure.
}
\label{fig:sc_integration}
\end{figure}

The service module has been sketched selecting some COTS cubesats components listed in
Table \ref{tab:service_module}, that will be implemented within  4 additional units. Therefore,
the complete configuration will be a 12U cubesat, $3\times 2 \times 2$.
All these elements, service module and payload can be integrated in the $12$U cubesat
as seen in Fig.~\ref{fig:sc_integration}.

\begin{table}
  \footnotesize
  \centering
  \caption{Service module components for EarthASAP.}
  \begin{tabular}{|c|c|c|c|c|}  
    \hline

    Unit name & Model and Supplier & Mass &
    Power  & Volume\\
                      &                                           & (g) & consumption & (mm$^3$)\\
    \hline
    Solar panel & {\scriptsize P110A GOMSpace (DK)} & 29 & 2300mW & $98\times 91.6\times 5.5$ \\
    Star Tracker & {\scriptsize St-400 Hyperion Technologies (NL)}& 280 & 700mW steady & $53.8\times 53.8\times 90.5$\\
    Reaction wheels &{\scriptsize RW400 Hyperion Technologies (NL)} & 340 & $<2    500$mW peak & $50\times 50\times 27.5$\\
Reaction Control system &{\scriptsize GOMSpace (DK)} & 220 & $2$W peak, $0.35$W standby & $100\times 100\times 30$\\
    On board computer & {\scriptsize CP400.85 Hyperion Technologies (NL)} &  7 &  $<1000mW$ peak & $20\times 50\times$ 10 \\
Telemetry Unit & {\scriptsize S band Transmitter }& 85 & $13$W transmitter & $99\times 93\times 15$ \\
     & {\scriptsize ISIS High Data Rate S-Band Transmitter   (NL) }& & & \\
    Battery &{\scriptsize  NanoPowerBPX GOMSpace (DK)} &  500 & &  $91.6\times 85.9\times 40$\\
    
    \hline
  \end{tabular}
  \label{tab:service_module}
 \end{table}

\section{Observation strategy and data analysis}\label{sec:observation_strategy}
EarthASAP will map the Earth, its exosphere and magnetosphere, the heliosphere,
and the lunar poles with a $20^o \times 30^o$ field imager and an angular
resolution of $3'$.\par

\begin{figure}
  \centering
  \includegraphics[width = 0.7\textwidth]{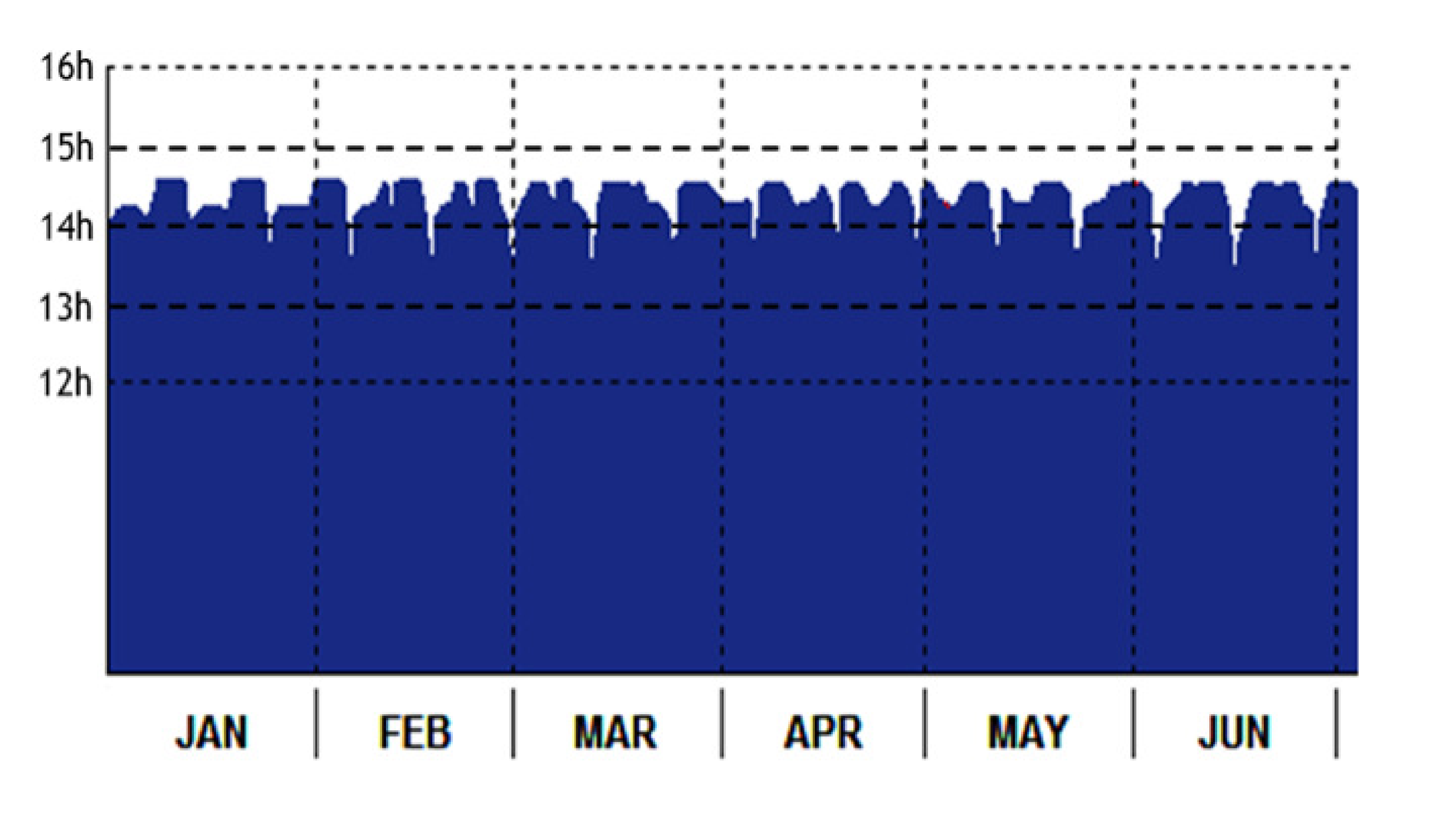}
  \caption{Estimated EarthASAP data relay Lunar Orbiter Visibility.}
  \label{fig:relay}
\end{figure}

The satellite  is designed for a low eccentricity lunar polar orbit, at 500km of the surface.
EarthASAP is designed to communicate with Earth via a Lunar Orbiter in $800 \time 8000$ km polar orbit. 
Simulating the EarthASAP and the data relay orbiter a daily coverage between 13.5 and 14.5 hours is obtained.
Figure \ref{fig:relay}  depicts the EarthASAP data relay Lunar Orbiter visibility.

\begin{figure}
  \centering
  \includegraphics[width = \textwidth]{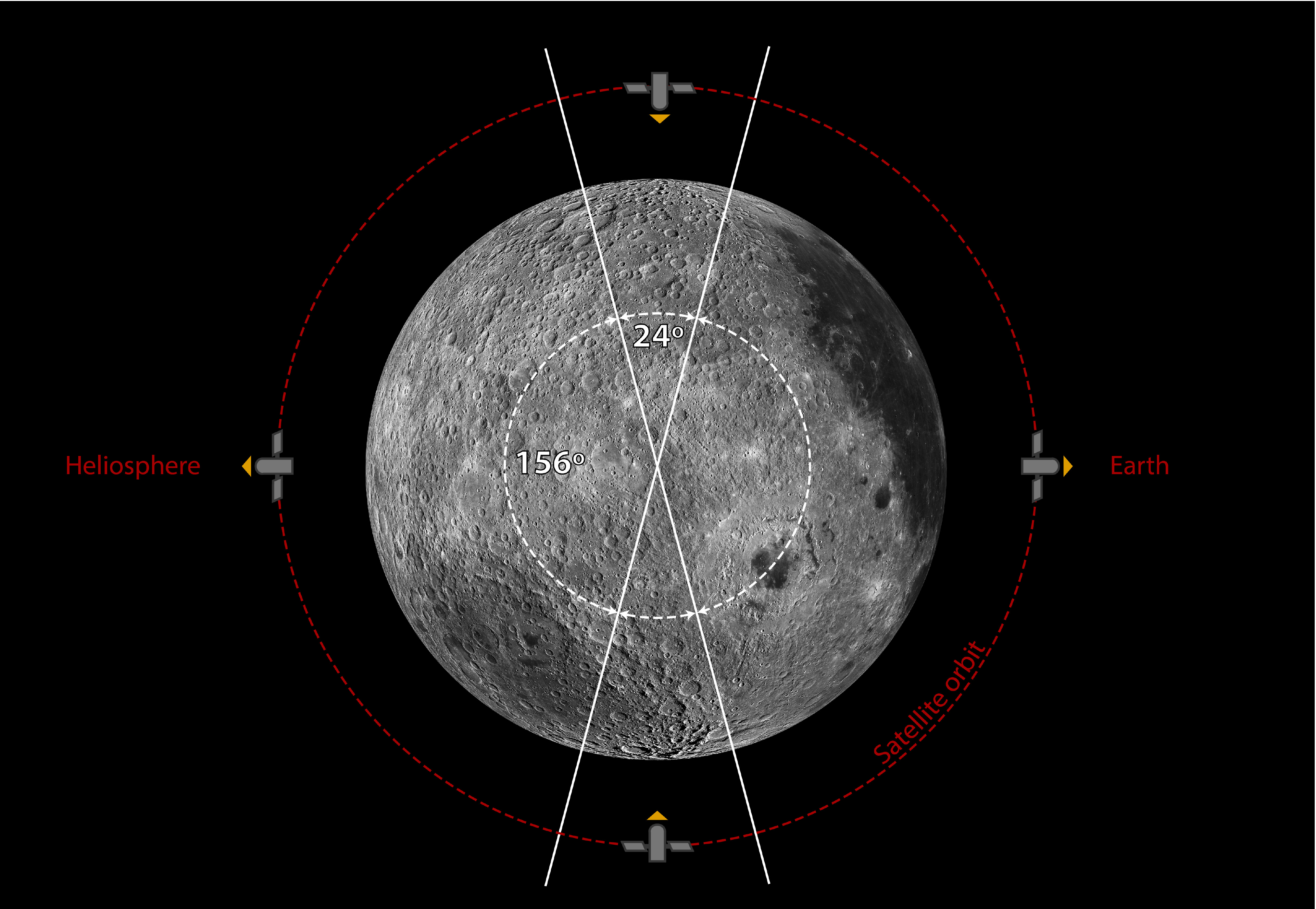}
  \caption{Mission planning for EarthASAP. While orbiting the Moon, the satellite
    will map systematically the Earth's exosphere, the heliosphere and the lunar poles with
  an optimized duty cycle.}
  \label{fig:orbit_sketch}
\end{figure}

The orbital period is 2.64 hours. For each orbital period, a basic cycle will be implemented consisting in surveying
during 64 minutes the Earth's exosphere, 12 minutes the South pole of the Moon, 64 minutes
the heliosphere and finally 12 minutes the North lunar pole. This duty cycle is sketched
in Figure \ref{fig:orbit_sketch}.

As the Orbiter is assumed to provide data relay services, only the
fuel to correct the orbit of the cubesat is required. The communications with the Lunar
Orbiter depends on the position in the orbit, but our simulations show that we
could obtain a daily coverage between 13.5h and 14.5h. \par

\begin{figure}
  \centering
  \includegraphics[width =0.7 \textwidth]{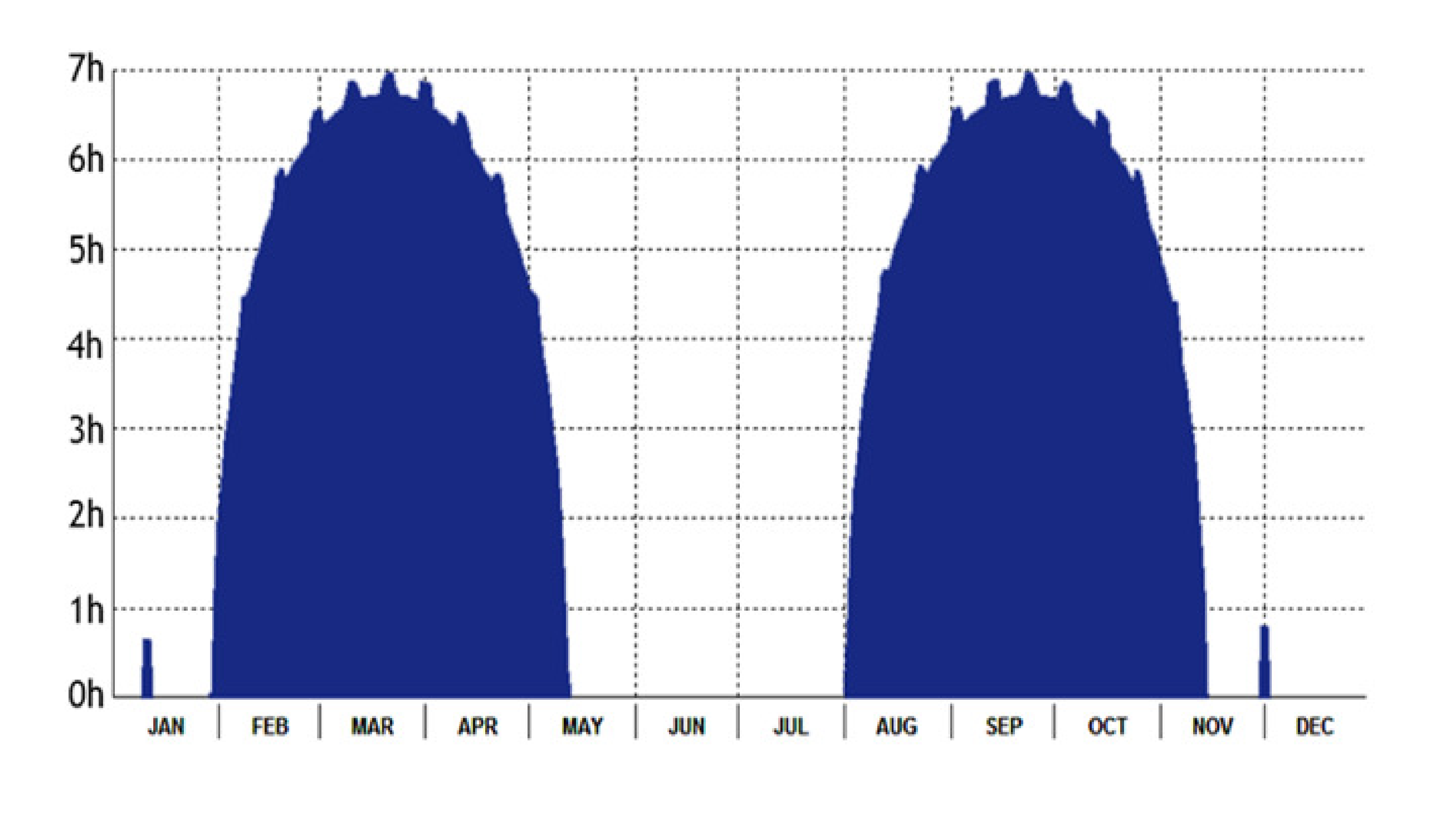}
  \caption{EarthASAP yearly eclipse evolution.}
  \label{fig:eclipse}
\end{figure}

The EarthASAP 500 km circular polar orbit will experiment every year two eclipse seasons. 
Each eclipse season will last about 3.5 months. The maximum eclipse duration will be about 
45 minutes. Statistically, EarthASAP will spend 13.1\% of the time in shadow. The mean eclipse 
duration is 36.6 minutes with a standard deviation of 9.3 minutes. In Figure \ref{fig:eclipse} 
the EarthASAP yearly eclipse evolution with the daily hours of shadows is depicted.

EarthASAP will produce images at a rate of 1 frame per minute in different
filters. This means around $\bf{6048}$ Mbytes/day. An on board processing 
with a compression factor $\bf{2}$ implies total amount of TM of roughly 
$\bf{3629}$ Mbytes/day. This must be downloaded during the averaged $\bf{14}$ hrs
when the relay spacecraft is visible, following Fig.~\ref{fig:relay}.
Presuming a protocol overhead of $\bf{20}$\%, the 
required bandwidth is around $\bf{72}$ Kbytes/sec.
This requirement can be fulfilled using
the S-band transmitter listed in Table~\ref{tab:service_module}.
This proposed S-band transmitter provides $\bf{537.5}$ Kbytes/sec for downlink,
which is enough for our purposes.

When defining the duty plan, one must take into 
account the available power budget, which will determine the 
activation of the different elements of the spacecraft.
For instance, we must consider the power consumption of the selected S-band transmitter,
or the eclipse seasons.
An initial power budget of the service module components
and payload is found in Table~\ref{tab:power_budget},
with a nominal consumption close to $\bf{15}$W, and a peak consumption
of $\bf{21}$W.

\begin{table}
\centering
\caption{EarthASAP power budget.}
\begin{tabular}{|c|c|c|c|}
\hline
\rule[-1ex]{0pt}{3.5ex}  Unit & Peak (W) & Duty Cycle (\%) & Average (W)\\
\hline\hline
\rule[-1ex]{0pt}{3.5ex}    Solar Panels (x9) & $20.7$ & $70$\% & $14.49$ \\    \hline
\rule[-1ex]{0pt}{3.5ex}    Star tracker & $-0.7$ & $100$\% & $-0.7$ \\    \hline
\rule[-1ex]{0pt}{3.5ex}    Reaction Wheels & $-2.5$ & $25$\% & $-0.6$ \\    \hline
\rule[-1ex]{0pt}{3.5ex}    Reaction Control System & $-2.0$ & (when needed) & $-0.35$ \\    \hline
\rule[-1ex]{0pt}{3.5ex}    On board Computer & $-1.0$ & $100$\% & $-1.0$ \\    \hline
\rule[-1ex]{0pt}{3.5ex}    TM/TC unit & $-13.0$ & $58$\% & $-7.6$ \\    \hline
\rule[-1ex]{0pt}{3.5ex}    Payload & $-4.0$ & $100$\% & $-4.0$ \\    \hline

\end{tabular}
\label{tab:power_budget}
\end{table}

The Reaction Control system (RCS) is needed for unloading the reaction wheels.
Periodic maintenance periods will be included in the duty plan to include these activities, 
when the power consumption may increase. 
Other periods that can reduce the continuous observations are the eclipses.
The batteries can provide just around $\bf{5.5}$hrs 
of full operations. Therefore, during eclipses, if the payload is kept is active, 
one may decide to interrupt the downlink of data.

During other periods, no image production will be possible. These periods
are, for instance, those when the Sun constraint could be violated because the
Sun enters in the FoV of the instrument when observing the Earth. 
This roughly happen during full moon phases, when no image acquisition is possible.
The duty plan can insert maintenance activities here.

\subsection{Data analysis and interpretation}\label{sec:data}
The contamination by cosmic rays will be removed and
geometrical distortions will be corrected on board. The optical system is designed to introduce a
well-known distortion that will be evaluated on ground tests and re-assessed during commissioning.
Flat fielding and an additional geometric correction to compensate for the variation
of the viewing angle will also be applied on board. Note that the angular
velocity of EarthASAP will be $\omega = 2.3^o$ min$^{-1}$, resulting in
a variation of the Earth projected size by a factor $\cos \omega t$, that will run
faster close to the poles. A final list with event centroids and time will
be generated and transferred to Earth for every pointing and filter during a
given orbit. \\

Ground based processing will include fitting and subtraction of the heliospheric
background. Lunar observations and comparison with previous missions
(mainly LRO) will be used for flat fielding and calibration. Scattered solar
radiation will be the main source for absolute flux calibration. \par

EarthASAP will take advantage of the wide field imaging to simultaneously obtain 
information on solar flux reaching the Earth, the ultraviolet background,
and the targets (Earth, Moon, Solar System) for most objects. Protocols will be
implemented to detect and track transient events such as solar flares reaching
the Earth, solar storms and geo-storms. \par

Finally, \lya~ maps will be used to build a 3D structure of the exosphere and determine
the transfer of \lya~ photons through it. The \lya~ source function will be calculated
directly from the data and made available to the community for future exo-Earth transit
calculations and data interpretation.

\section{Summary}

This article summarizes the basic scientific and technological characteristics of a 12U
cubesat to be set in Lunar orbit. It is shown that the small 8U payload is enough to address 
important problems concerning the nature of gas escape from the Earth's atmosphere (and Earth-like 
planets). The small EarthASAP mission is also well suited to monitor the 
hydration of the rocks in the lunar poles, and the heliosphere. The observatory can be manufactured
in few years, at a cost below five million euros.

Up to now, three technical problems have been identified but all can be addressed with the
existing technology. First, the proposed optical design
is subjected to a moderate amount of barrel distortion. However, this effect can be
corrected by the post-processing of the images. Second, the proposed off-axis
telescope requires a lateral input window and a deployable baffle. These
two elements might be placed in the side of the payload unit but require designing
a structure that allows the window to maintain the structural properties of the
system. Also, a baffle or lid  would be needed to close the input windows
during launch and transit phases. Third, in order to align correctly the off-axis
mirror system and maintain it independently of temperature changes, it is
fundamental to use materials with small thermal expansion coefficient such as 
carbon fibre based composites or inbar.

Most important, to make the Moon accessible to EarthASAP and, in general, to lunar cubesat 
missions, it is required  that a communications relay is set close to the Moon. This is the only means
to keep their power needs within the cubesat scale.

\acknowledgments 
This article has been partially funded by the Ministerio de Economia y Competitividad of Spain under grant, ESP2015-68908-R
and ESP2017-87813-R. 


\bibliography{references}   
\bibliographystyle{spiejour}   

\end{spacing}
\end{document}